\documentclass[journal]{IEEEtran}

\usepackage{graphicx}
\usepackage{multirow}
\usepackage{lipsum}
\usepackage{tcolorbox}
\usepackage{bm}
\usepackage{xr}
\usepackage{float}
\usepackage{amsmath}
\usepackage{amssymb}
\usepackage{amsthm}
\usepackage{amsfonts}
\usepackage{booktabs,siunitx}
\usepackage{xcolor}
\usepackage{soul}
\usepackage{cleveref}
\usepackage{svg}
\usepackage[normalem]{ulem}
\usepackage{xspace}
\usepackage{algorithm}
\usepackage[noend]{algpseudocode}

\usepackage{amsthm}

\theoremstyle{definition}
\newtheorem{definition}{Definition}[section]

\theoremstyle{definition}
\newtheorem{problem}{Problem}

\begin{document}

\title{Motifs in temporal hypergraphs}

\author{Quintino Francesco Lotito, Lorenzo Betti, Federico Battiston, and Giuseppe Francesco Italiano%
\thanks{Q. F. Lotito and L. Betti are with the Department of Network and Data Science, Central European University, Vienna, Austria (e-mail: lotitoq@ceu.edu; betti\_lorenzo@phd.ceu.edu).}%
\thanks{F. Battiston is with the Center for Networks and Complexity, Luiss University, Rome, Italy; the Department of AI, Data and Decision Sciences, Luiss University, Rome, Italy; and the Department of Network and Data Science, Central European University, Vienna, Austria (e-mail: fbattiston@luiss.it).}%
\thanks{G. F. Italiano is with the Department of AI, Data and Decision Sciences, Luiss University, Rome, Italy (e-mail: gitaliano@luiss.it).}%
}

\maketitle

\begin{abstract}
Network motifs, recurrent local patterns of interactions in graphs, provide fundamental insights on the interplay between structure and functionality in complex systems. Many real-world systems are not well represented by traditional static pairwise networks, as interactions may involve groups of nodes, occur over time, or encode directionality. In this paper, we introduce temporal motifs for hypergraphs and directed hypergraphs, extending motif analysis to timestamped many-body interactions. We formalize the corresponding mining problem, study the combinatorial structure of these motifs, and develop exact algorithms for their enumeration. In particular, we propose a dynamic programming algorithm that substantially reduces the computational cost of motif mining, achieving orders of magnitude speedups on empirical datasets. We also introduce a null model for temporal hypergraphs to assess the statistical over- and under-expression of motifs. Applying the proposed framework to real-world datasets from different domains, including face-to-face contacts, scientific collaborations, e-mail exchanges, and Bitcoin transactions, we show that temporal hypergraph motifs reveal distinct forms of local organization across systems. Finally, we demonstrate their use as an exploratory tool through focused case studies on persistent patterns in scientific collaborations and e-mail communications.
\end{abstract}


\section{Introduction}
\label{sec:introduction}

Real-world networked systems can be described and differentiated by the recurring patterns of interactions among their constituent units. These patterns, called \emph{network motifs}, have long been studied in the literature, as their frequency is thought to affect system dynamics~\cite{milo2002network,milo2004superfamilies, schwarze2020motifs} and they are therefore often referred to as the \emph{building blocks} of complex systems~\cite{milo2002network}. Besides their foundational role, motifs have found application in several downstream tasks in different domains, such as revealing recurrent gene-regulatory patterns shared by organisms from bacteria to humans~\cite{shen2002network, alon2007network}, extracting fingerprints of social networks~\cite{hong2014social, juszczyszyn2008local}, detecting early warning signs of structural change in financial networks~\cite{saracco2016detecting}, and studying direct and indirect species interactions in ecological systems~\cite{bascompte2009assembly, simmons2019motifs}.

Driven by the growing availability of new empirical data, networks have been extended in order to accommodate more complex interaction features, allowing a more faithful representation of real-world systems. In fact, representing real-world complex systems with simple, static, and pairwise edges may hide important aspects of their underlying structural organization. This also changes the kind of local patterns one may want to study. In a \emph{hypergraph}, which is a natural framework for studying many-body interactions, a motif is not simply a small arrangement of edges, but a configuration that may include \emph{hyperedges} of different sizes and different forms of overlap~\cite{lotito2022higher, lee2020hypergraph}. In a temporal network, in which interactions take place over time, patterns of links together with their temporal ordering may carry important information~\cite{paranjape2017motifs, kovanen2011temporal}. 

These richer representations call for motif definitions that account for how interactions of different sizes are organized across time, potentially in settings where such interactions may also encode direction. In this work, we propose definitions, mining algorithms, and statistical tools for temporal motifs in hypergraphs and directed hypergraphs. We analyze empirical datasets with timestamped many-body interactions, considering both undirected hyperedges, which represent the group interactions, and directed hyperedges, in which interactions encode a source set and a target set of nodes. Our results show that temporal hypergraph motifs distinguish different forms of local organization across domains. In face-to-face contacts, over-expression is associated with rapid group formation and repeated sub-group encounters, whereas in scientific collaborations it is associated with persistent author cores and gradual team reconfiguration. In directed hypergraphs, over-expressed motifs reveal further forms of organization. In e-mail exchanges, they capture short and dynamic communication threads, with a rich landscape of broadcasts, replies, and relay-like behavior. In Bitcoin transactions, they highlight repeated transaction templates in which source and target roles remain more stable over time. These analyses provide a statistical characterization of motif expression across domains, while two focused use cases show how temporal motifs can yield descriptors of group persistence and individual participation.

The main contributions of this work can be summarized as follows:

\begin{itemize}
    \item We extend the notion of motifs to hypergraphs and directed hypergraphs with timestamped interactions, study their combinatorial structure, and formalize the corresponding mining problem.
    \item We develop a baseline mining algorithm and a more efficient algorithm based on dynamic programming, reducing execution times on real-world datasets by several orders of magnitude.
    \item We introduce a null model for temporal hypergraphs and use it to characterize the over- and under-expression of motifs across empirical datasets from different domains.
    \item We demonstrate the use of temporal motifs as an exploratory tool through two exemplary case studies on persistent patterns in scientific collaborations and e-mail communications.
\end{itemize}

The rest of the paper is structured as follows. \Cref{sec:related} reviews related work on generalized network models and motif analysis. \Cref{sec:preliminaries} introduces the main definitions and formalizes the mining problem. \Cref{sec:combinatorics} studies the combinatorial space of temporal hypergraph motifs, while \Cref{sec:algorithms} presents the exact algorithms used for motif counting. \Cref{sec:null-model} describes the null model used to evaluate motif enrichment. Finally, \Cref{sec:experiments} reports the empirical results, \Cref{sec:applications} presents the practical use cases, and \Cref{sec:conclusion} lists some concluding remarks.

\section{Related work}
\label{sec:related}

In this section, we first review recent advances in the modeling of real-world complex systems, with particular emphasis on temporal and higher-order networks. We then discuss how the framework of network motifs has evolved to address the additional complexity introduced by these generalized network models.

\subsection{Temporal and higher-order interactions in complex systems}
Over the last few decades, networks have established themselves as the standard language to describe and study systems of interacting units. Traditional approaches describe these systems in terms of static and pairwise links between nodes~\cite{boccaletti2006complex, cimini2019statistical}. While useful, this representation showed drawbacks in encoding important structural information of empirical systems, where interactions naturally evolve over time or involve groups of more than two units at the same time.

Many-body -- higher-order -- interactions appear in many contexts, including social systems~\cite{patania2017shape, cencetti2021temporal, iacopini2024temporal}, ecological communities~\cite{grilli2017higher}, chemical reactions~\cite{jost2019hypergraph} and metabolic pathways~\cite{traversa2023robustness}, as well as brain networks~\cite{petri2014homological, santoro2023higher}. Hypergraphs~\cite{berge1973graphs} are a natural way to encode these interactions, since hyperedges can connect any number of nodes simultaneously, and provide a richer description of systems whose structure cannot be fully reduced to pairwise connections~\cite{battiston2020networks, battiston2021physics}. This perspective has motivated the development and extension of several tools to characterize higher-order organization, including measures of centrality~\cite{benson2019three, tudisco2021node}, community structure~\cite{eriksson2021choosing, contisciani2022inference, ruggeri2023community}.

When group interactions are timestamped, systems are naturally described through
temporal hypergraphs, where each hyperedge carries an associated timestamp.
This framework has been used to study a variety of evolving systems, such as human face-to-face interactions~\cite{iacopini2024temporal} and scientific collaborations~\cite{chowdhary2024team,lerner2023micro}.
By incorporating the temporal dimension, it becomes possible to account for the dynamic nature of group interactions and study the mechanisms underpinning the evolution of group interactions.
Higher-order temporal interactions have been shown to share characteristics with their pairwise counterparts, such as bursty activity~\cite{cencetti2021temporal}, while also exhibiting distinctive features related to their multi-body nature, such as the mechanisms by which groups form, grow, and dissolve across time~\cite{iacopini2024temporal}. The notion of temporal neighborhood has also been extended to higher-order interactions, providing a way to characterize the time evolution of social interactions from the perspective of egocentric structures~\cite{arregui2024patterns}.
Finally, other methods analyze temporal hypergraphs in terms of multi-scale structural properties~\cite{mancastroppa2024structural} and temporal correlations~\cite{gallo2024higher}.

\subsection{Network motifs beyond simple graphs}
Motif analysis is a popular framework for describing and characterizing complex networks at their local scale.
Following their introduction in static graphs, network motifs have been extended in several directions to account for increasingly rich representations of empirical systems. These extensions relax different assumptions of the original framework, including the absence of weights or temporal information, the existence of a single interaction layer, or the binary nature of links. For example, Onnela et al. extended motifs to weighted networks by describing subgraphs through their intensity and coherence~\cite{onnela2005intensity}. Battiston et al. extended motif analysis to multilayer networks by studying subgraphs that appear across multiple layers more often than expected~\cite{battiston2017multilayer}, a problem also closely related to the isomorphism problem in the same network framework~\cite{kivela2018isomorphism}. In temporal networks, motifs are used to capture patterns of interactions that occur within limited time windows~\cite{paranjape2017motifs,kovanen2011temporal}.  

More recently, motif analysis has also been extended to hypergraphs. Lee et al. studied and characterized the local structure of real-world hypergraphs in terms of recurrent overlapping patterns among triples of hyperedges of arbitrary size~\cite{lee2020hypergraph}, also incorporating their temporal dimension~\cite{lee2021thyme+}. More specifically, the temporal patterns introduced in THyMe+ describe the arrival order and the emptiness of the seven Venn-diagram regions associated with exactly three connected hyperedges, independently of the total number of participating nodes. Lotito et al. instead considered a direct extension to hypergraphs of the seminal notion of network motifs introduced by Milo et al.~\cite{milo2002network}, extracting over-expressed subhypergraph patterns induced by small sets of typically three and four nodes~\cite{lotito2022higher}, or up to five nodes in the approximated setting~\cite{lotito2024exact}. This framework has been further extended to directed hypergraphs, enabling the analysis of hyperedges that encode directional flow~\cite{lotito2026microscale}. In this work, we introduce the temporal dimension into this node-based framework for both undirected and directed hypergraphs, characterizing how these patterns are born, evolve, and die over time.

%%%%%%% Preliminary
\section{Preliminaries and problem statement}
\label{sec:preliminaries}
Here, we first introduce basic definitions and then formalize the mining problem of our interest.

\begin{figure*}
    \centering
    \includegraphics[width=0.9\linewidth]{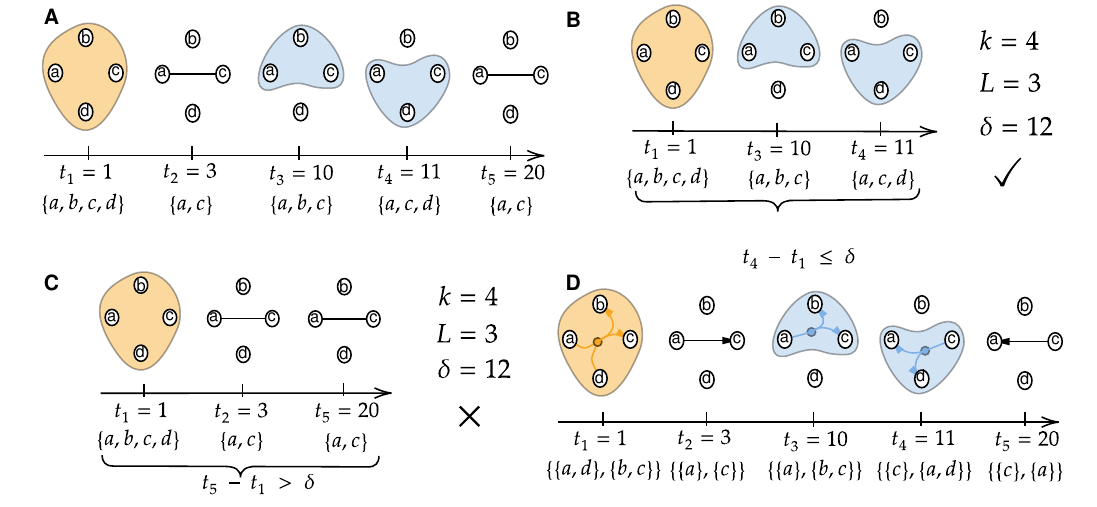}
    \caption{\textbf{Schematic representation of motif discovery in temporal hypergraphs.} A) A temporal hypergraph with $4$ nodes and $5$ timestamped hyperedges. Links are straight dark lines. Hyperedges are colored shadow areas, with different colors indicating different interaction sizes. B) Example of a valid motif extracted from such a temporal hypergraph. The motif is composed of $K=4$ nodes and $L=3$ hyperedges. The difference between the latest and the earliest hyperedge in the pattern is less than the constraint $\delta = 12$. C) Example of a pattern that does not count as a motif, in our definition, due to failing to meet the $\delta = 12$ constraint. D) We can represent the directionality of the information flow of the hyperedges by encoding a source and a target set of nodes.} 
    \label{fig:motif_schema}
\end{figure*}

\begin{definition}
A \emph{temporal hypergraph} is a pair $H=(V,E)$ where $V$ is a set of nodes and
$
E = \{(\tau_1,e_1), (\tau_2,e_2), \dots, (\tau_m,e_m)\}$
is a set of time-stamped hyperedges. Each $\tau_i$ is a timestamp and each $e_i$
is either
(i) an undirected hyperedge $e_i\subseteq V$ with $|e_i|\ge 2$, or
(ii) a directed hyperedge $e_i=(S_i,T_i)$ with $S_i,T_i\subseteq V$ non-empty and
$S_i\cap T_i=\emptyset$, where $S_i$ and $T_i$ encode the source and target nodes of interaction $i$. For a directed hyperedge $e_i=(S_i,T_i)$, we define its size as $|e_i|:=|S_i\cup T_i|=|S_i|+|T_i|$.
The \emph{static projection} of $H$ is the (static) hypergraph
$H_{\mathrm{static}}=(V,E_{\mathrm{static}})$ obtained by discarding timestamps and collapsing repeated hyperedges, i.e.,
$E_{\mathrm{static}}=\{e_1,\dots,e_m\}$.
\end{definition}

\begin{definition}
A \emph{$K$-node, $L$-hyperedge, $\delta$-temporal motif} is a sequence
$
M = \big( (\tau_1, \varepsilon_1), (\tau_2, \varepsilon_2), \dots, (\tau_L, \varepsilon_L) \big)
$
with $\tau_1 < \tau_2 < \dots < \tau_L$ and $\tau_L - \tau_1 \le \delta$.
The hyperedges in $M$ involve exactly $K$ distinct nodes, and their static projection is connected. To focus on the role of higher-order interactions, we also require at least one genuine higher-order interaction in the pattern, i.e., $\exists j \in \{1,\dots,L\}
\quad \text{such that} \quad
|\varepsilon_j| \ge 3$.
\end{definition}

Having defined a temporal motif, we next specify when it occurs in an observed temporal hypergraph.

\begin{definition}
An \emph{occurrence} of $M$ in $H$ is a choice of $L$ events
$(\hat\tau_1,\hat e_1),\dots,(\hat\tau_L,\hat e_L)\in E$
and an injective mapping $\phi: V(M)\to V$ such that:
(i) $\hat\tau_1<\dots<\hat\tau_L$ and $\hat\tau_L-\hat\tau_1\le\delta$;
(ii) for each $j\in\{1,\dots,L\}$, $\hat e_j = \phi(\varepsilon_j)$, where $\phi$ is applied element-wise to node sets and, in the directed case,
$\phi((S,T))=(\phi(S),\phi(T))$).
\end{definition}

Motifs that differ only by a relabelling of their nodes represent the same pattern.

\begin{definition}
  Two temporal motifs
  $M=\big((\tau_1,\varepsilon_1),\dots,(\tau_L,\varepsilon_L)\big)$
  and
  $M'=\big((\tau'_1,\varepsilon'_1),\dots,(\tau'_L,\varepsilon'_L)\big)$
  are \emph{isomorphic} if there exists a bijection
  $\psi:V(M)\to V(M')$ such that
 $\psi(\varepsilon_j)=\varepsilon'_j$ for every
  $j\in\{1,\dots,L\}$, with source and target sets preserved in the directed
  case.
  \end{definition}

\begin{problem}
Given a temporal hypergraph $H=(V,E)$ and parameters $(K,L,\delta)$, compute the number of
occurrences in each isomorphism class of $(K,L,\delta)$-temporal motifs in $H$.
\end{problem}

In Fig.~\ref{fig:motif_schema}, we provide a schematic summary of the basic definitions introduced here, including an example of a temporal directed and undirected hypergraph, and valid and invalid instances of motifs.  

\section{Combinatorial analysis}
\label{sec:combinatorics}
Before discussing the technical algorithmic aspects related to the mining problem, here we study the size of the combinatorial space of temporal motifs. While the proposed bounds are not intended to be tight, they still provide a useful characterization of the rate at which the temporal motif space grows.

We denote with $m_T(K,L)$ the number of
non-isomorphic \emph{connected} temporal motifs on $K$ nodes and temporal length $L$ (i.e., sequences
of $L$ hyperedges). Throughout this section, we assume $K\ge3$, fix a labelled vertex set $V = \{1,\dots,K\}$ and consider length-$L$
sequences of time-stamped hyperedges on $V$. Let $\mathcal{H}(K)$ denote the size of the catalogue
of distinct static hyperedges on $K$ labelled vertices (excluding the empty set and singleton
hyperedges). The total number of labelled temporal sequences of length $L$ is then
\[
\mathcal{H}(K)^L.
\]

Each non-isomorphic motif on $K$ nodes can be represented by several different labelled sequences:
any permutation of the node labels yields another labelled instance of the same structural pattern.
In the worst case, a motif can generate $K!$ distinct labelled sequences, one for each permutation
of $\{1,\dots,K\}$. Conversely, any two labelled sequences that differ only by a permutation of the
labels correspond to the same structural motif. In particular, each non-isomorphic motif on $K$
nodes corresponds to \emph{at most} $K!$ labelled sequences.

On one hand, $m_T(K,L)$ cannot exceed the total number of labeled sequences, so
\[
m_T(K,L) \;\le\; \mathcal{H}(K)^L.
\]

To focus on connected motifs, let $L_{\mathrm{conn}}$ denote the number of labelled sequences
whose static projection on $V$ is connected and that contain at least one higher-order interaction. Since each connected motif contributes at most $K!$
labelled sequences, we have
\[
m_T(K,L) \;\ge\; \frac{L_{\mathrm{conn}}}{K!}.
\]

To derive a lower bound for $L_{\mathrm{conn}}$, we can enforce connectivity by requiring that the temporal sequence contains a spanning hyperedge
$e^\star$ (i.e., an interaction that involves all $K$ nodes) at least once. In the undirected case
we may take $e^\star = V$, while in the directed case we may take $e^\star = (S^\star,T^\star)$ with
$S^\star \cup T^\star = V$ (and $S^\star,T^\star$ disjoint and non-empty). Any sequence that includes
$e^\star$ has a connected static projection (and involves all $K$ nodes). The number of labelled sequences that contain $e^\star$ at least once
is obtained by subtracting those that avoid $e^\star$ entirely:
\[
L_{\mathrm{conn}}
\;\ge\;
\mathcal{H}(K)^L - \big(\mathcal{H}(K)-1\big)^L.
\]
Consequently,
\[
m_T(K,L)
\;\ge\;
\frac{\mathcal{H}(K)^L - \big(\mathcal{H}(K)-1\big)^L}{K!}.
\]

Combining the two bounds, we obtain
\begin{equation}
\label{eq:general_bounds}
\frac{\mathcal{H}(K)^L - \big(\mathcal{H}(K)-1\big)^L}{K!}
\;\le\;
m_T(K,L)
\;\le\;
\mathcal{H}(K)^L.
\end{equation}
Here, the division by $K!$ in the lower bound accounts for permutations of the $K$ vertex labels.

We now instantiate this general framework for specific hypergraph models by substituting the
corresponding catalogue sizes into Eq.~\eqref{eq:general_bounds}.

For the \textbf{undirected case}~\cite{lotito2022higher}, the catalogue of valid hyperedges on
$K$ vertices is
\[
\mathcal{H}(K) = 2^K - K - 1.
\]

In the \textbf{directed case}~\cite{lotito2026microscale}, interactions are defined by disjoint
non-empty source and target node sets. The number of possible directed hyperedges on $K$ labelled
vertices is
\[
\mathcal{H}_{\rightarrow}(K) = 3^{K} - 2^{K+1} + 1.
\]
Substituting $\mathcal{H}(K)$ and $\mathcal{H}_{\rightarrow}(K)$ into Eq.~\eqref{eq:general_bounds}
gives the corresponding bounds for undirected and directed temporal motifs, respectively.

\section{Algorithms}
\label{sec:algorithms}
We now describe the two exact algorithms used to count temporal motifs in the empirical hypergraphs. Both algorithms share the same first step: they construct a reduced set of candidate node supports, and then count motif occurrences only inside those supports.

\subsection{Common support computation}
Instead of searching over arbitrary node sets, we use the seed-hyperedge strategy introduced for static hypergraph motifs~\cite{lotito2024exact}. We call the set of $K$ nodes involved in a motif occurrence its \emph{support}. Candidate supports are generated from observed higher-order hyperedges, which we call \emph{seeds}. Here, we focus on $K\in\{3,4\}$, corresponding to the motif sizes considered in our empirical analysis. A seed of size $K$ already defines a full support, while a smaller seed is extended through an overlapping event whenever their union contains exactly $K$ nodes. This construction is exact for the motif sizes considered here, because every valid motif contains at least one higher-order interaction. For larger values of $K$, the same strategy can be generalized by recursively extending a seed through overlapping events. Algorithm~\ref{alg:supports} summarizes the support computation shared by the baseline and LocalWindow-DP.

\begin{algorithm}
\caption{\textsc{BuildSupports}$(H,K)$}
\label{alg:supports}
\begin{algorithmic}[1]
\Require Temporal hypergraph $H=(V,E)$, motif size $K\in\{3,4\}$
\Ensure Candidate supports $\mathcal{S}_K$
\State Let $V(e)$ denote the node set of event $e$; in the directed case, $V(e)=S_e\cup T_e$
\State Initialize $\mathcal{S}_K\gets\emptyset$
\ForAll{events $e_i\in E$ with $3\leq |V(e_i)|\leq K$}
    \If{$|V(e_i)|=K$}
        \State Add $V(e_i)$ to $\mathcal{S}_K$
    \Else
        \ForAll{events $e_j\in E$ such that $V(e_i)\cap V(e_j)\neq\emptyset$}
            \If{$|V(e_i)\cup V(e_j)|=K$}
                \State Add $V(e_i)\cup V(e_j)$ to $\mathcal{S}_K$
            \EndIf
        \EndFor
    \EndIf
\EndFor
\State \Return $\mathcal{S}_K$
\end{algorithmic}
\end{algorithm}

After this shared step, the two algorithms differ in how they count motif occurrences. The baseline enumerates temporal sequences directly, whereas LocalWindow-DP uses dynamic programming to reuse partial sequences within the same time window.

\subsection{Baseline}
The baseline algorithm is the direct exact enumeration procedure. It first computes $\mathcal{S}_K$ with Algorithm~\ref{alg:supports}. For each candidate support $S\in\mathcal{S}_K$, it extracts the local event sequence $E_S$, formed by all events whose nodes are contained in $S$. It then enumerates length-$L$ subsequences with strictly increasing timestamps, extending a partial sequence only while it remains within the time window $\delta$. A completed sequence is counted only if it uses all $K$ nodes of the support, contains at least one higher-order interaction, and has connected static projection. Valid sequences are canonically relabelled and added to the corresponding motif count.

This algorithm is useful as an exact reference, but it is not efficient. Its cost is dominated by the enumeration of length-$L$ subsequences inside each candidate support. Even after restricting the search to candidate supports, dense local event sequences can generate many candidates. A useful proxy for this cost is $\sum_{S\in\mathcal{S}_K} N_L(E_S,\delta)$, where
$N_L(E_S,\delta)$ is the number of length-$L$ subsequences of the local event sequence $E_S$ with
strictly increasing timestamps whose first and last timestamps differ by at most $\delta$. In a dense local window with $a$ events, this number can be as large as $\binom{a}{L}$.

\begin{algorithm}
\caption{\textsc{BaselineEnumeration}$(H,K,L,\delta)$}
\label{alg:baseline}
\begin{algorithmic}[1]
\Require Temporal hypergraph $H=(V,E)$, motif size $K$, motif length $L$, time window $\delta$
\Ensure Motif counts $C$
\State Sort events $E=\{(t_i,e_i)\}_{i=1}^{M}$ by nondecreasing timestamp
\State $\mathcal{S}_K\gets\textsc{BuildSupports}(H,K)$
\State Initialize $C[m]\gets 0$ for all motif labels $m$
\ForAll{$S\in\mathcal{S}_K$}
    \State $E_S\gets\{(t_i,e_i)\in E: V(e_i)\subseteq S\}$, preserving temporal order
    \State Relabel nodes in $S$ locally as $\{1,\dots,K\}$
    \ForAll{length-$L$ subsequences $Q$ of $E_S$ with $t_{i_1}<\dots<t_{i_L}$ and $t_{i_L}-t_{i_1}\leq\delta$}
        \If{$Q$ uses all $K$ nodes, contains a higher-order event, and has connected static projection}
            \State $m\gets\mathrm{CanonicalLabel}(Q)$
            \State $C[m]\gets C[m]+1$
        \EndIf
    \EndFor
\EndFor
\State \Return $C$
\end{algorithmic}
\end{algorithm}

\subsection{Local window dynamic programming}
LocalWindow-DP also starts from the supports computed by Algorithm~\ref{alg:supports}, but replaces explicit subsequence enumeration with prefix counting. For a fixed support $S$, each event in $E_S$ is mapped to a local state on the labelled nodes of $S$. In the undirected case, a state is a node subset of size at least two. In the directed case, a state is a pair of disjoint non-empty source and target subsets. We denote the resulting local state space by $\Omega_K$, and precompute which length-$L$ state sequences are valid motifs, together with their canonical labels.

The algorithm scans each local sequence $E_S$ once. During the scan, it maintains a sliding window of events whose timestamps are within $\delta$ of the current event. Inside this window, sparse dynamic programming tables $\mathrm{dp}_r$ store how many active prefixes of length $r$ have each local state sequence, for $r=1,\dots,L-1$. Events sharing the same timestamp are
processed as a batch. When a new batch arrives, motif counts are updated by matching the state $x_i$ of each event against valid prefixes of length $L-1$ formed at earlier timestamps. After evaluating motif counts for the entire batch, we update the prefix tables in decreasing order of length, processing all events in the batch at each length before proceeding to the next. Therefore, simultaneous events may
represent alternative choices in an occurrence, but cannot appear together in the same occurrence.

The left boundary of the window is handled exactly. When an event leaves the window, the prefix counts that use that specific event as their first element are subtracted from the corresponding tables. This preserves the same temporal constraint as the baseline enumeration, but replaces repeated subsequence generation with incremental prefix counting. Algorithm~\ref{alg:localwindowdp} summarizes the resulting LocalWindow-DP procedure.

LocalWindow-DP is exact and returns the same counts as the baseline. By avoiding repeated work on overlapping local subsequences, it becomes substantially faster than direct enumeration on the empirical datasets considered. After the shared support computation, a useful proxy for the counting cost is $\sum_{S\in\mathcal{S}_K}|E_S|$, since each local event sequence is scanned once. The multiplicative factors depend only on $K$ and $L$, through the finite sets of local states and prefix transitions. In our experiments, $K,L\in\{3,4\}$, so these factors are constant.

\begin{algorithm}
\caption{\textsc{LocalWindow-DP}$(H,K,L,\delta)$}
\label{alg:localwindowdp}
\begin{algorithmic}[1]
\Require Temporal hypergraph $H=(V,E)$, motif size $K$, motif length $L$, time window $\delta$
\Ensure Motif counts $C$
\State Sort events $E=\{(t_i,e_i)\}_{i=1}^{M}$ by nondecreasing timestamp
\State $\mathcal{S}_K\gets\textsc{BuildSupports}(H,K)$
\State Build the local state space $\Omega_K$ on $K$ labelled nodes
\State Precompute valid state sequences $\mathcal{A}_K\subseteq\Omega_K^L$ and labels $\ell(a)$
\State Initialize $C[m]\gets 0$ for all motif labels $m$
\ForAll{$S\in\mathcal{S}_K$}
    \State $E_S\gets\{(t_i,e_i)\in E: V(e_i)\subseteq S\}$, preserving temporal order
    \State Relabel nodes in $S$ locally as $\{1,\dots,K\}$ and convert $E_S$ into states $(t_i,x_i)$
    \State Initialize sparse maps $\mathrm{dp}_r:\Omega_K^r\to\mathbb{N}$ for $r=1,\dots,L-1$ and an active queue $W$
    \ForAll{timestamp batches $B_t=\{(t_i,x_i)\in E_S:t_i=t\}$ in increasing $t$}
        \While{$W\neq\emptyset$ and $t-\mathrm{time}(\mathrm{front}(W))>\delta$}
            \State $(t_0,x_0)\gets\mathrm{pop\_front}(W)$
            \State Subtract from $\mathrm{dp}_1,\dots,\mathrm{dp}_{L-1}$ prefixes starting with $(t_0,x_0)$
        \EndWhile
        \ForAll{$(t,x_i)\in B_t$}
            \ForAll{prefixes $p\in\Omega_K^{L-1}$ such that $(p,x_i)\in\mathcal{A}_K$}
                \State $C[\ell(p,x_i)]\gets C[\ell(p,x_i)] + \mathrm{dp}_{L-1}[p]$
            \EndFor
        \EndFor
        \For{$r=L-2,L-3,\dots,1$}
            \ForAll{$(t,x_i)\in B_t$}
                \ForAll{$p\in\mathrm{supp}(\mathrm{dp}_r)$}
                    \State $\mathrm{dp}_{r+1}[(p,x_i)]\gets \mathrm{dp}_{r+1}[(p,x_i)] + \mathrm{dp}_r[p]$
                \EndFor
            \EndFor
        \EndFor
        \ForAll{$(t,x_i)\in B_t$}
            \State $\mathrm{dp}_1[(x_i)]\gets \mathrm{dp}_1[(x_i)] + 1$
            \State $\mathrm{push\_back}(W,(t,x_i))$
        \EndFor
    \EndFor
\EndFor
\State \Return $C$
\end{algorithmic}
\end{algorithm}

\section{Null models and evaluation of temporal motifs}
\label{sec:null-model}
An important step in motif analysis is evaluating the statistical relevance of the observed frequencies of the patterns. A common way of performing this task is to compare empirical raw counts against the counts observed in suitable randomizations of the empirical data, extracting the z-scores associated with each pattern. Let $\mathcal{M}$ be the set of unlabelled motifs. For each $m\in\mathcal{M}$ let $C_m^{\mathrm{obs}}$ denote the count observed in the empirical data. We generate $R$ independent realizations of the null model, and write $C_m^{(1)},\dots,C_m^{(R)}$ for the corresponding null counts of motif $m$. For each motif, we compute the null mean $
\mu_m=\frac{1}{R}\sum_{r=1}^R C_m^{(r)},
$
the null standard deviation
$
\sigma_m=\sqrt{\frac{1}{R}\sum_{r=1}^R\left(C_m^{(r)}-\mu_m\right)^2},
$
and the associated $z$-score
$
z_m=\frac{C_m^{\mathrm{obs}}-\mu_m}{\sigma_m}.
$
Thus, $z_m$ provides a standardized measure of the deviation between the empirical count $C_m^{\mathrm{obs}}$ and the distribution produced by the
null ensemble. We use these scores to rank motifs within each dataset. Motifs with large positive or negative $z$-scores are respectively referred to as \emph{over-} or \emph{under-expressed} relative to the null model.

Since temporal motifs are identified within a 
time window $\delta$, we should design the null model taking into account this temporal constraint. A simple random permutation of timestamps over the whole dataset would strongly disrupt the empirical event-rate profile -- the extent to which related events occur close in time. In particular, it could make motif occurrences artificially rare by spreading events too sparsely in time relative to the window $\delta$~\cite{kovanen2011temporal}. Therefore, the null model should randomize the temporal ordering of interactions while preserving local event density at a coarse temporal scale.
Taking inspiration from~\cite{tsvetkova2016dynamics}, we use a windowed timestamp-shuffling method. This null leaves hyperedge participants unchanged and randomizes only their assignment to timestamps, within a temporal scale and hyperedge size constraint. In particular, an undirected event $(t,e)$ retains the same hyperedge $e$, while a directed event $(t,(S,T))$ retains the same source set $S$ and target set $T$. As a result, the static hypergraph is unchanged. The randomization is carried out within coarse time windows of width $\Delta$. Let $t_0$ be the minimum timestamp in the data, and define the coarse window index by $b(t)=\lfloor (t-t_0)/\Delta \rfloor$. Events are grouped by the pair consisting of their window index $b(t)$ and their \emph{event signature}. For undirected hypergraphs, the signature is the event size $|e|$; for directed hypergraphs, it is the ordered pair $(|S|,|T|)$. Within each group, the observed timestamps are uniformly permuted and then reassigned to the original event participants. This construction keeps the structure unchanged by shuffling timestamps only across hyperedges of the same size (or hyperedges with the same source and target sets size), destroying fine-scale temporal correlations. By choosing $\Delta$ similar to the motif window $\delta$, the null model perturbs the temporal ordering of events while avoiding the excessive temporal sparsification that would arise from an unconstrained timestamp shuffle. It is therefore well-suited for testing whether an observed motif profile reflects temporal ordering effects beyond static higher-order structure.

\begin{algorithm}
\caption{\textsc{WindowedTimestampShuffle}$(E,\Delta)$}
\label{alg:null}
\begin{algorithmic}[1]
\Require Time-stamped events $E=\{(t,e)\}$ or $E=\{(t,(S,T))\}$, null window size $\Delta$
\Ensure Shuffled event list $\widetilde{E}$ with preserved event signatures
\State $t_0 \gets \min\{t : (t,\cdot)\in E\}$
\State Initialize map $\mathcal{G}$ from grouping keys to event indices
\ForAll{event indices $i$ with event $(t_i,e_i)$ or $(t_i,(S_i,T_i))$}
    \State $w_i \gets \left\lfloor (t_i-t_0)/\Delta \right\rfloor$
    \State $\sigma_i \gets (|e_i|)$ in the undirected case, or $\sigma_i \gets (|S_i|,|T_i|)$ in the directed case
    \State Append $i$ to $\mathcal{G}[(w_i,\sigma_i)]$
\EndFor
\State Initialize $\widetilde{t}_i \gets t_i$ for all events
\ForAll{groups $I \in \mathcal{G}$}
    \State $T_I \gets [t_i : i\in I]$
    \State Uniformly permute $T_I$
    \ForAll{paired values $(i,\tau)$ with $i\in I$ and $\tau\in T_I$}
        \State $\widetilde{t}_i \gets \tau$
    \EndFor
\EndFor
\State $\widetilde{E} \gets \{(\widetilde{t}_i,e_i)\}$ or $\{(\widetilde{t}_i,(S_i,T_i))\}$
\State Sort $\widetilde{E}$ by nondecreasing time
\State \Return $\widetilde{E}$
\end{algorithmic}
\end{algorithm}

\section{Experimental results}
\label{sec:experiments}
We evaluate temporal hypergraph motifs on four empirical datasets, publicly available through Hypergraphx-data~\cite{lotito2026hypergraphx}. Two datasets are undirected: face-to-face contacts and scientific collaborations. The other two are directed: e-mail exchanges and Bitcoin transactions. Face-to-face contacts are extracted from the SocioPatterns project, and represent daily interactions between students of different classes in a high school over $5$ days~\cite{mastrandrea2015contact}. Scientific collaborations have been extracted from a dump of ArXiv, and represent preprints connecting the different authors in the field of Computer Science from 1998 to 2022. E-mail exchanges have been collected from the Enron project. Finally, the Bitcoin dataset has been built considering the first $200$k transactions among different wallets during the month of November 2014. The datasets also differ in temporal resolution: face-to-face contacts have a granularity of $20$ seconds, scientific collaborations are indexed at daily resolution, and both e-mail exchanges and
Bitcoin transactions use timestamps in seconds. Table~\ref{tab:dataset-summary} reports the main dataset statistics. 

For the runtime evaluation, we extract temporal motifs with $K\in\{3,4\}$ nodes and $L\in\{3,4\}$ temporal hyperedges. For the statistical analysis, we focus on $K\in\{3,4\}$ and $L=3$, use $R=10$ null-model realizations, and rank motifs by their z-scores within each dataset. Here, we use z-scores descriptively to rank motif enrichment, rather than for formal hypothesis testing. We refer again to Table~\ref{tab:dataset-summary} for the experimental details, where the motif parameter $\delta$ and the null-model parameter $\Delta$ are set equal within each system.

\begin{table*}[!t]
\centering
\caption{Summary statistics of the empirical temporal hypergraphs.}
\label{tab:dataset-summary}
\small
\begin{tabular}{lccccccc}
\hline
Dataset & Directed & $|V|$ & $|E|$ & $|E_2|$ & $|E_3|$ & $|E_4|$ & $\delta=\Delta$\\
\hline
Face-to-face contacts & $\times$ & $327$ & $130{,}133$ & $99{,}127$ & $20{,}433$ & $6{,}628$ & $1$h\\
Scientific collaborations & $\times$ & $402{,}361$ & $378{,}829$ & $78{,}328$ & $92{,}171$ & $71{,}516$ & $5$y\\
E-mail exchange & $\checkmark$ & $84{,}172$ & $235{,}395$ & $138{,}811$ & $25{,}773$ & $13{,}847$ & $1$d\\
Bitcoin transactions & $\checkmark$ & $597{,}291$ & $200{,}000$ & $21{,}122$ & $81{,}165$ & $45{,}646$ & $1$d\\
\hline
\end{tabular}
\end{table*}

\subsection{Performance evaluation}

The experiments have been carried out on a machine with an 8-core (2.2GHz) Intel Xeon CPU and 94GB of RAM, running Ubuntu 20.04.4 LTS. The algorithms are implemented in Python3 and available on Hypergraphx~\cite{lotito2023hypergraphx}.

\begin{figure*}
    \centering
    \includegraphics[width=\linewidth]{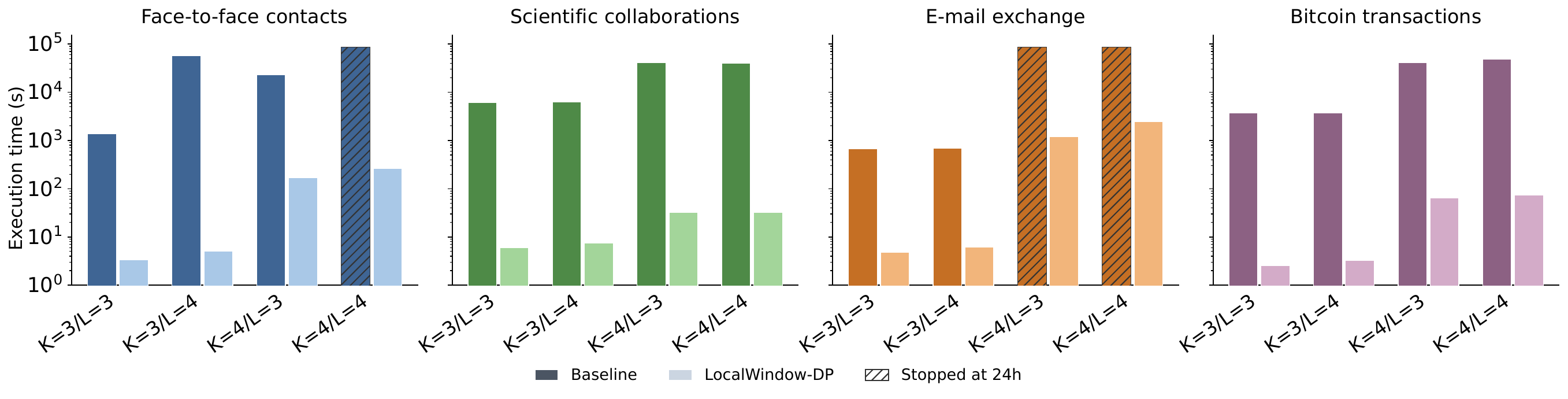}
    \caption{\textbf{Runtime comparison of exact temporal motif-counting algorithms.} Mean wall-clock execution time over 10 runs for the baseline support enumeration and LocalWindow-DP across the four empirical datasets. Each panel corresponds to one dataset, and bars report the runtime for the four $(K,L)$ configurations considered in the experiments. The $y$-axis is logarithmic. Hatched bars denote configurations stopped after $24$ hours and plotted at the timeout value.}
    \label{fig:runtime}
\end{figure*}

In Fig.~\ref{fig:runtime}, we show the execution runtime in seconds for the baseline and the dynamic programming approach on the four datasets and on multiple $K,L$ configurations. Each bar reports the mean wall-clock execution time over $10$ independent runs. The proposed DP approach consistently outperforms the baseline, often by several orders of magnitude, making it feasible to study configurations that would otherwise be impractical due to high runtime. Variability in execution time across runs of the same configuration was negligible relative to these differences and is therefore omitted from the logarithmic-scale figure.

\subsection{Undirected hypergraphs}
\begin{figure*}
    \centering
    \includegraphics[width=0.9\linewidth]{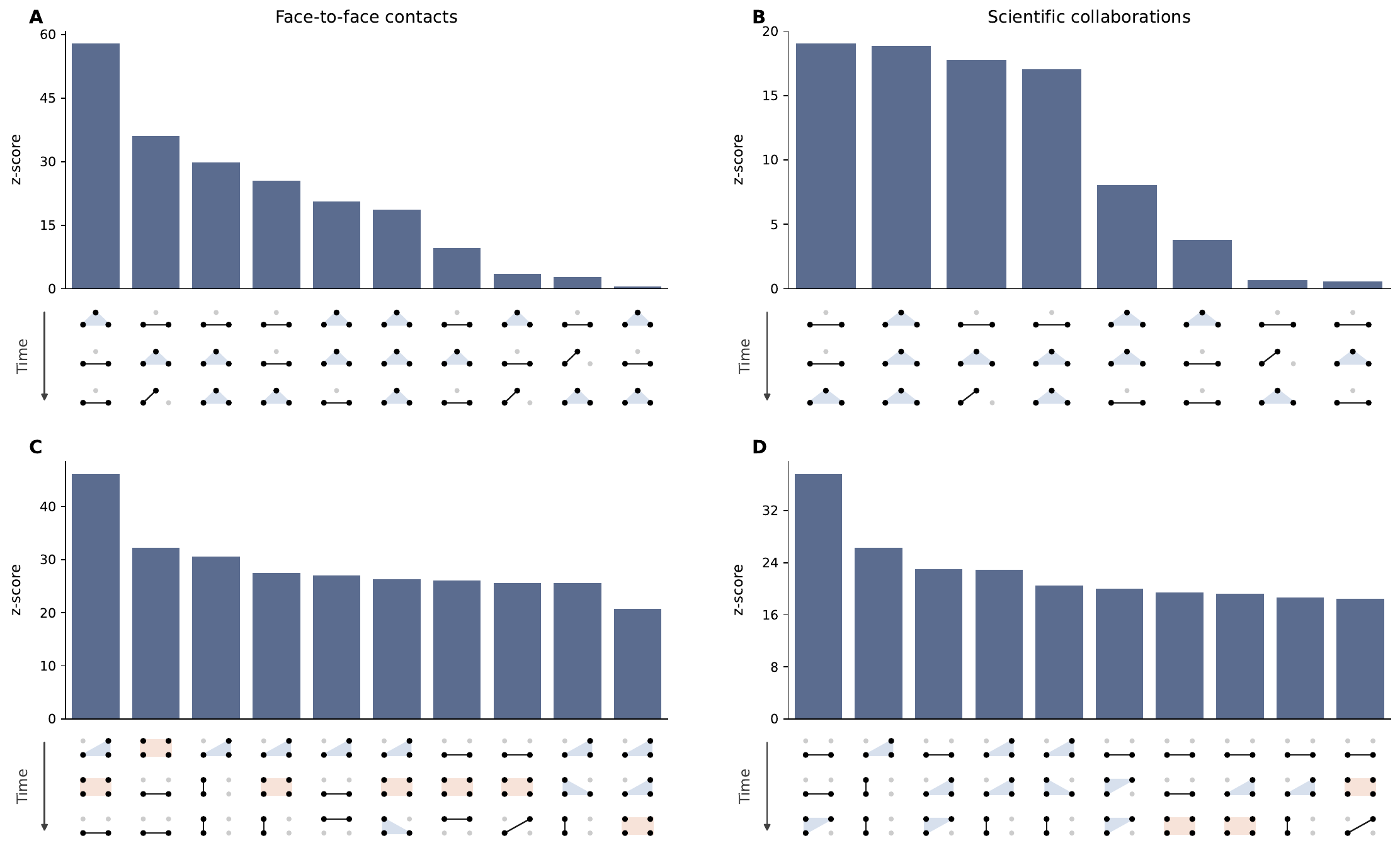}
    \caption{\textbf{Over-expressed motifs in real-world temporal hypergraphs.} Each panel reports the motifs with the largest positive null-model z-scores for one dataset and one motif size, highlighting temporal patterns that occur more often in the data than expected under the timestamp-shuffling null model. The two columns correspond to face-to-face contacts and scientific collaborations, and the two rows to $3$-node and $4$-node motifs. Graphlets below the bars show the temporal ordering of the hyperedges. Time flows downward inside each motif.}
    \label{fig:overexpressed-motifs-undirected}
\end{figure*}

We first consider simple, undirected temporal hypergraphs. Figure~\ref{fig:overexpressed-motifs-undirected} shows the most over-expressed motifs of order $3$ and $4$ in face-to-face contacts and scientific collaborations. In these systems, motifs describe how group interactions are born, split, disappear, and reappear over time, often around stable cores and changing peripheral participants.

The top over-expressed face-to-face motifs describe the short-time life cycle of small interaction groups. Several patterns contain a group encounter that is followed by smaller overlapping contacts, or by a later reappearance of part of the same group. Other motifs show the opposite direction: dyadic or partial contacts are followed by a larger group interaction. Their over-expression suggests that groups in face-to-face interactions are not formed as isolated events. They are born from smaller contacts, split into overlapping subgroups, and sometimes reappear or partially reappear shortly after.

Scientific collaborations show other interesting forms of interaction patterns. The enriched motifs often contain a persistent subset of authors that appears across multiple papers, while other authors enter or leave the group. In this case, the motif is naturally interpreted as a collaboration core with a changing periphery. This is consistent with team-based scientific production, where a small group of authors keeps working together across related papers, while other collaborators join or leave over time.

\subsection{Directed hypergraphs}
\begin{figure*}
    \centering
    \includegraphics[width=0.9\linewidth]{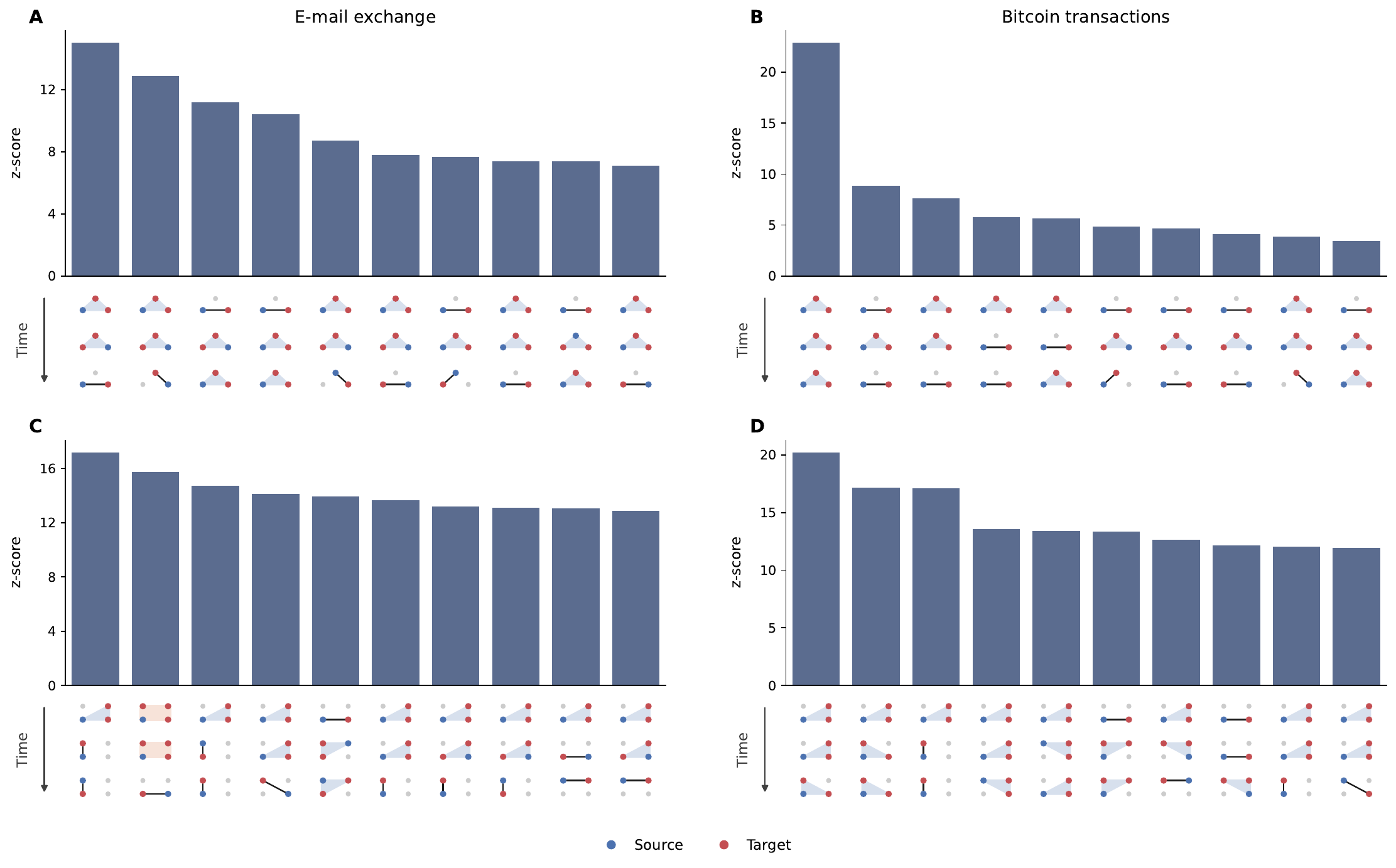}
    \caption{\textbf{Over-expressed motifs in real-world temporal directed hypergraphs.} Each panel shows the motifs with the highest positive z-scores relative to the null model, thus identifying enriched directed higher-order interaction patterns. Columns correspond to e-mail exchange and Bitcoin transactions, while rows correspond to $3$-node and $4$-node motifs. Graphlets display the temporal ordering of events, with blue nodes denoting sources and red nodes denoting targets.}
    \label{fig:overexpressed-motifs-directed}
\end{figure*}

We now turn our attention to directed hypergraphs with timestamped hyperedges. We recall that, in this case, each interaction has a source set and a target set that encode information flow.  Figure~\ref{fig:overexpressed-motifs-directed} shows the most over-expressed motifs of order $3$ and $4$ in e-mail and Bitcoin data.

The top over-expressed e-mail motifs contain repeated group communication together with local role changes. Several patterns include broadcast-like
events, where one sender reaches multiple recipients, either repeated over time or followed by narrower exchanges with part of the same audience. Others contain a reply-like structure, where a previous target later sends back to a previous sender. We also observe relay-like patterns, where a previous target becomes a sender toward another node. Their over-expression suggests that these modules are not isolated noise, but form recurring local threads in which a group audience is created, reused, and partially redirected over time. 

The Bitcoin motifs highlight a different organization, suggesting less role alternation than in e-mail exchanges and a stronger persistence of transaction roles. In the top over-expressed patterns, many motifs reuse the same source-to-target configuration, or keep one address in a stable source role while the target set changes only partially. Other patterns suggest redistribution-like routines, where one source address repeatedly sends to partially overlapping target sets. Their enrichment is consistent with repeated payments or other common practices in which source and target roles remain comparatively stable.

\section{Practical use cases}
\label{sec:applications}
Beyond motif counting and statistical analysis, temporal motifs can be used as building blocks for application-driven analyses. 
Interestingly, a motif is not only a local temporal pattern to be detected, but also a compact unit from which one can derive interpretable observables tailored to the system under study. 
Below, we provide two descriptive case studies on scientific collaborations and e-mail communications.
Here, the focus shifts from identifying over-expressed motif types to using all extracted motif occurrences to characterize persistent groups and individual participation profiles.

Before turning to the applications, we formalize the notion of \emph{core} of a temporal motif, which provides a common way to characterize persistent participation in both use-case domains.

\begin{definition}[Core of a temporal motif]
Let $M=\big((\tau_1,\varepsilon_1),\dots,(\tau_L,\varepsilon_L)\big)$ be an occurrence of a temporal motif. Its \emph{core} is the set of nodes that appear in every interaction of the occurrence, namely $\mathrm{core}(M)=\bigcap_{i=1}^{L}\varepsilon_i$ in the undirected case, and $\mathrm{core}(M)=\bigcap_{i=1}^{L}(S_i\cup T_i)$ in the directed case, where $\varepsilon_i=(S_i,T_i)$. Thus, in the directed setting, the core records persistent participation regardless of whether a node appears in the source or in the target set.
\end{definition}

Beyond identifying persistent groups within motifs, we also consider how often individual nodes belong to these cores and how their interaction partners change over time. 
To characterize these complementary aspects of individual participation, we define \emph{coreness} as the fraction of a node's motif participations in which it belongs to the core. We also define \emph{neighbor turnover} as the mean Jaccard distance between neighborhoods at consecutive active timestamps separated by at most $\delta$.

\subsection{Persistent collaboration cores in scientific production}
In scientific collaborations, a natural question is whether repeated local interaction patterns are organized around stable nuclei of authors, or whether they are mainly driven by continual turnovers in group composition. Temporal motifs provide a convenient tool for extracting these insights. In particular, the previously introduced notion of core identifies the subset of authors that remains present across all interactions of a motif occurrence, and therefore isolates the stable nucleus around which a local collaboration pattern unfolds.

In this application, we use $K=4$, $L=3$ motifs with $\delta=1826$ days, and analyze the Computer Science arXiv data from 1998 to 2022. Persistence is evaluated across five disjoint $5$-year periods, from 1998--2002 to 2018--2022, extracting motifs separately within each period. For the author-level panels, we retain authors active in at least three periods and assign roles from coreness. For an author with $n$ active periods, we average coreness over the first and last $\lfloor n/2\rfloor$ periods to obtain earlier and later values, respectively. Roles are labelled as \emph{peripheral} for averages below $0.25$, \emph{core} for averages above $0.75$, and \emph{mixed} otherwise. Panel~D further requires at least $10$ core memberships and $10$ eligible neighborhood transitions.

\begin{figure*}
    \centering
    \includegraphics[width=0.9\linewidth]{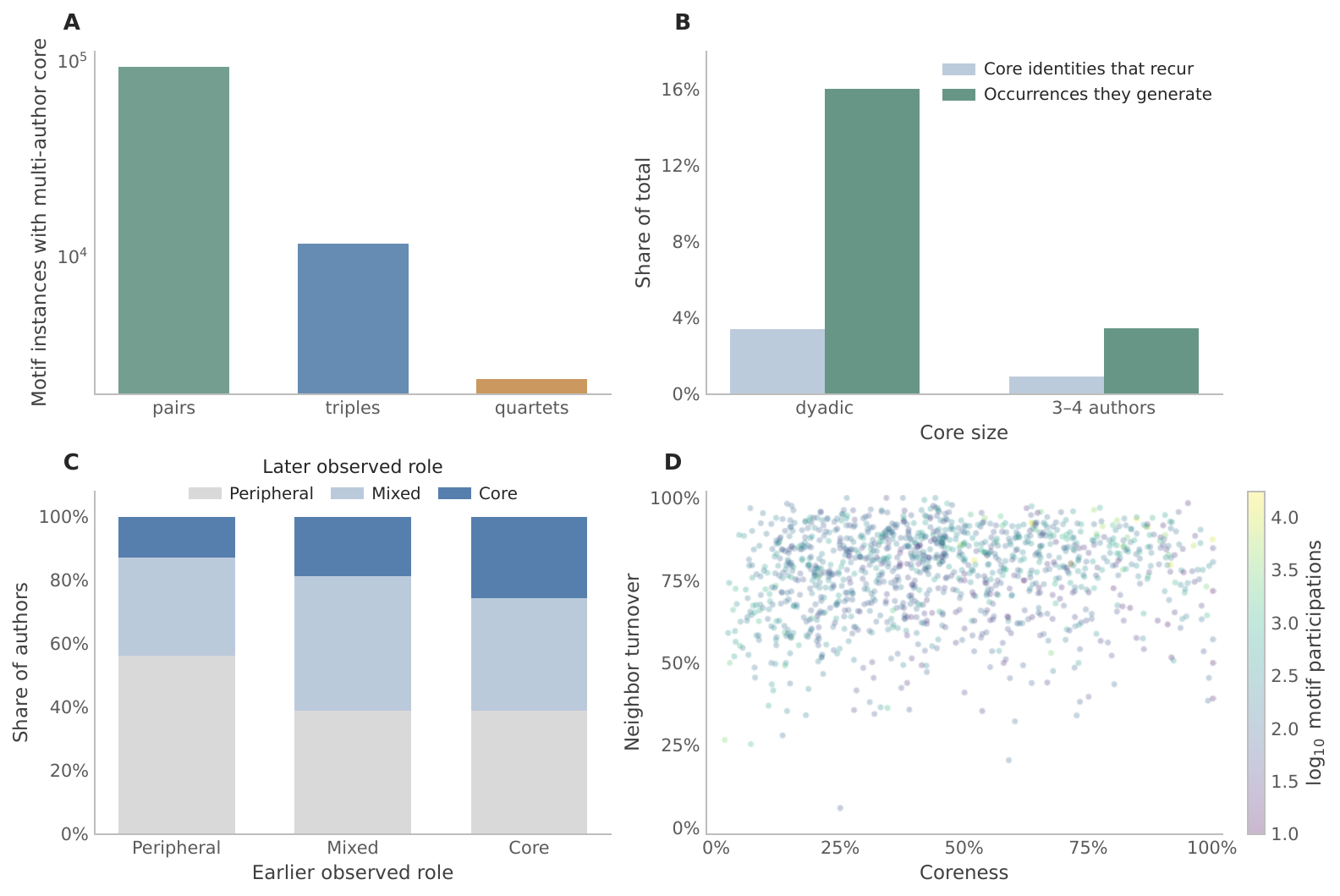}
    \caption{\textbf{Cores in temporal motifs reveal higher-order collaboration dynamics.} A) Total motif occurrences with multi-author cores, stratified into pairs, triples, and quartets. B) Shares of core identities recurring across at least two disjoint periods and of motif occurrences associated with these identities, within each core-size category. C) Later observed roles conditional on earlier roles for authors, classified as peripheral, mixed, or core from their coreness. D) Author cartography in which the $x$-axis is coreness, the $y$-axis is neighbor turnover, and color encodes $\log_{10}$ motif participations.}
    \label{fig:coauth-application}
\end{figure*}

Figure~\ref{fig:coauth-application} shows that the distinction between cores and peripherals is informative. Panel~A reports motif occurrences with multi-author cores, stratified by core size. The dominant contribution within this subset comes from pairs of authors. Panel~B complements this picture by measuring recurrence across temporal periods: a core is recurrent if the same exact author set appears as a core in at least two periods. The bars compare the share of recurrent core identities with the share of motif occurrences associated with them, within each core-size category. Interestingly, a small set of recurring author groups accounts for a disproportionate share of motif occurrences. Panels~C and~D shift the focus from motifs to single authors. Panel~C summarizes changes in authors' roles between earlier and later active periods, distinguishing peripheral, mixed, and core roles. While most authors initially classified as peripheral remain so, most authors initially classified as core are later classified as mixed or peripheral, highlighting an asymmetry in the persistence of these roles. Finally, panel~D combines coreness with neighbor turnover. The cartography reveals heterogeneous author profiles spanning almost the full coreness range, from predominantly peripheral participants to authors who consistently belong to motif cores. Neighbor turnover is generally high across this range, although authors with similar coreness can differ substantially in the stability of their collaboration partners.

Overall, temporal motifs provide a microscopic description of scientific collaborations that connects the recurrence of specific author groups with changes in individual participation. The analysis highlights the predominance of pairs among multi-author cores, the disproportionate contribution of recurring groups to motif occurrences, and heterogeneous author profiles characterized by changes in core participation and generally high neighbor turnover.

\subsection{Persistent communication cores in e-mail exchanges}
In communication data, local interaction patterns are heavily characterized by whether nodes switch between sending and receiving, and by how quickly one interaction is reciprocated by another. Directed temporal motifs provide a natural unit for studying communication groups together with the local organization of information flow.

In this application, we use the full Enron dataset and extract directed motifs with $K\in\{3,4\}$, $L=3$, and $\delta=1$ day. The node-level cartography in panel~C pools both motif sizes and is restricted to nodes with at least $10$ core memberships and $10$ eligible neighborhood transitions.

First, we formalize role changes within temporal motifs and weak reciprocity between their interactions.

\begin{definition}[Role change and weak reciprocity]
Let $M$ be an occurrence of a directed temporal motif, and let $v\in M$. Node $v$ undergoes a \emph{role change} within $M$ if there exist $i,j\in\{1,\dots,L\}$ such that $v\in S_i$ and $v\in T_j$, and is \emph{role-stable} otherwise. Following the role-based notion of reciprocity introduced for directed temporal hypergraphs in~\cite{lotito2026microscale}, two interactions $e_i=(S_i,T_i)$ and $e_j=(S_j,T_j)$ with $i<j$ exhibit \emph{weak reciprocity} if
\[
(T_i\cap S_j\neq\varnothing)\wedge(S_i\cap T_j\neq\varnothing).
\]
Equivalently, at least one node moves from target to source and at least one node moves from source to target across the two interactions. A directed motif occurrence displays \emph{weak reciprocity} if it contains at least one such pair.
\end{definition}

\begin{figure*}
    \centering    \includegraphics[width=\linewidth]{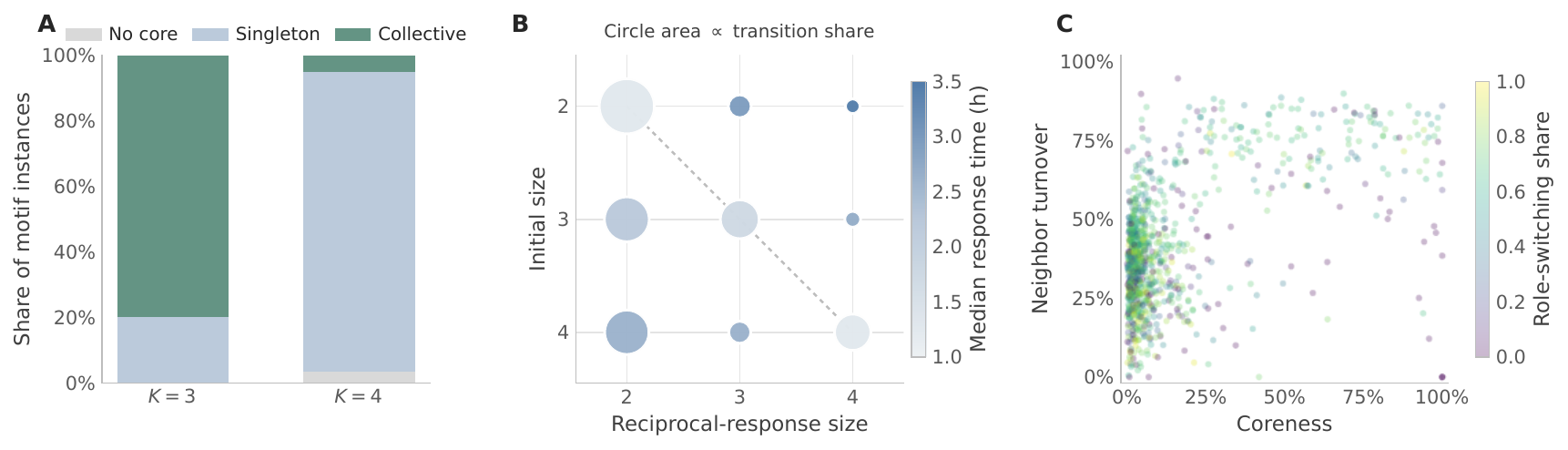}
    \caption{\textbf{Temporal motifs reveal reciprocity patterns and flow organization in e-mail exchanges.} A) $3$- and $4$-node motif occurrences stratified into empty, singleton, and collective cores, with collective cores containing at least two nodes. B) Initial versus earliest qualifying reciprocal-event size, defined as $|S|+|T|$. Circle area represents the transition share within each initial-size row, and color encodes median delay in hours, conditional on reciprocation within extracted motifs. C) Node-level cartography in which the $x$-axis measures coreness, the $y$-axis measures neighbor turnover, and color encodes role-switching share.}
    \label{fig:enron-application}
\end{figure*}

Figure~\ref{fig:enron-application} combines the analysis of motif cores and reciprocity in e-mail exchanges. Panel~A stratifies motif occurrences into empty, singleton, and collective cores. Persistent communication groups dominate $3$-node motifs, whereas $4$-node motifs are predominantly organized around a single persistent participant. Panel~B focuses on weak reciprocity and measures the size and delay of reciprocal interactions inside motifs. We pool reciprocal event pairs across both motif sizes, count each pair only once, and retain the earliest qualifying reciprocation for each initiating event. We consider only reciprocations within extracted motifs and the one-day horizon. The transition shares show that group interactions are frequently reciprocated through dyadic exchanges, indicating that reciprocation often involves only part of the original group. For each initial interaction size, median reciprocation delays are shorter when the reciprocal interaction has the same size. Panel~C moves to the node level, combining coreness with neighbor turnover. Color encodes role-switching share, the fraction of a node's core memberships in which it appears as both source and target within the motif. Many low-coreness nodes have relatively stable neighborhoods, suggesting repeated interactions with similar groups despite rarely participating in every interaction of a motif. These predominantly peripheral participants can nevertheless alternate between sending and receiving when they belong to a motif core. In contrast, high-coreness nodes often combine persistent participation with changing communication groups.

Altogether, this analysis suggests that e-mail communication combines persistent participants with others whose involvement is intermittent within local interaction patterns. Group interactions are often reciprocated through more selective exchanges, while persistent participants can interact with changing groups and frequently switch between sending and receiving.

\section{Conclusion}
\label{sec:conclusion}
In this work, we went beyond analyzing static patterns of hyperedges, introducing the notion of time. We proposed new definitions for motifs in temporal directed and undirected hypergraphs, we studied the combinatorics of such objects, we developed exact algorithms to extract them from real-world datasets and a null model for temporal hypergraphs to evaluate their over- and under-expression, allowing us to interpret the building blocks of temporal higher-order networks. Our dynamic programming approach achieves orders of magnitude speedups over direct enumeration on empirical datasets. Finally, two descriptive case studies show how motif-based observables characterize recurring collaboration cores and changes in author participation in scientific production, as well as persistent communication cores, reciprocity, and sender--receiver role changes in e-mail exchanges.

Given the complexity of the mining problem, interesting avenues for future work include the development of sampling algorithms able to trade off solution quality for performance~\cite{lotito2024exact}. Compared to the static problem, both the motif size $K$ and the temporal length $L$ can potentially affect the algorithmic landscape of the solving strategies, leading to multiple specialized solutions for the different cases.

Taken together, motifs in temporal hypergraphs provide a local-scale description of complex systems with higher-order interactions, capturing how higher-order interaction patterns unfold over time.

\section*{Acknowledgment}
This work was supported by the Italian Ministry of University and Research (MUR) under Decreto Direttoriale n. 1236 of 1 August 2023 – Bando FIS 2, project CODYBE (FIS-2023-02989), CUP B53C25003310001, admitted to funding by D.D. Prot. n. I.0018353 of 19 November 2025 and by the Austrian Science Fund (FWF) under project 10.55776/PAT1052824.

\appendix
\section{Appendix A: Anti-motifs}
\begin{figure*}
    \centering
    \includegraphics[width=0.9\linewidth]{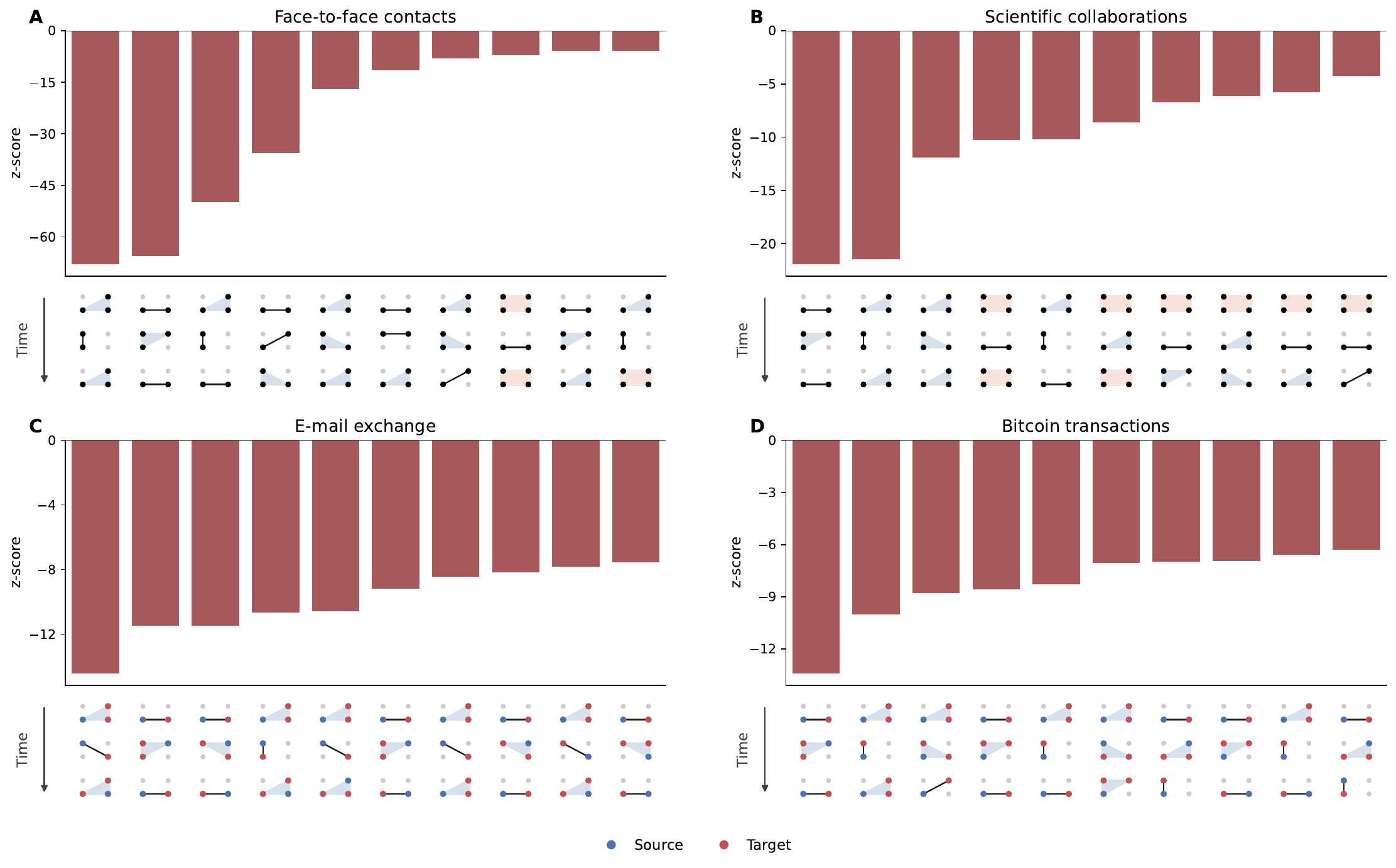}
    \caption{\textbf{Under-expressed motifs in real-world temporal hypergraphs.} Each panel reports the $K=4$, $L=3$ motifs with the most negative null-model z-scores. The first row corresponds to undirected datasets, while the second row corresponds to directed datasets.}
    \label{fig:anti-motifs}
\end{figure*}

In addition to over-expressed motifs, validating empirical frequencies against our null model allows the identification of \emph{anti-motifs}, i.e., temporal patterns that occur less often than expected after preserving the static hypergraph and the coarse temporal activity profile. Figure~\ref{fig:anti-motifs} reports the most under-expressed $K=4$, $L=3$ motifs. These patterns should not be interpreted as impossible configurations, but as local temporal arrangements that the empirical systems tend to suppress.

In the undirected datasets, several under-expressed motifs display fragmented patterns in which a group interaction is followed by weakly overlapping or disconnected subgroups. This complements the over-expression results in the main text. Face-to-face contacts and scientific collaborations do not only favor repeated cores and partially reappearing groups, but also suppress temporal arrangements in which the same four-node support is activated through poorly coherent group changes.

In the directed datasets, the under-expressed motifs often mix role changes with unstable source--target configurations. For e-mail, this means that not every possible reply or relay structure is equally likely: the data favor local threads with recognizable target reuse, while suppressing patterns where source and target roles change without a coherent communication context. For Bitcoin, most under-expressed motifs are consistent with a preference for stable transaction roles: transaction motifs tend to preserve stable roles, and patterns that combine repeated transactions with abrupt role reversals are comparatively avoided. 

\bibliographystyle{IEEEtran}
\bibliography{biblio}

\end{document}